# Nuclear shadowing in deep inelastic scattering on nuclei at low and medium $Q^2$

Edgar V. Bugaev
*Institute for Nuclear Research of the Russian Academy of Sciences, Moscow 117312, Russia*

Boris V. Mangazeev
*Department of Physics, Irkutsk State University, Irkutsk 664003, Russia*

Nuclear shadowing corrections to the structure functions of deep inelastic scattering of intermediate-mass nuclei are calculated at very low values of Bjorken $x$ and small values of $Q^2$ $\left(Q^2 \leq 5\,GeV^2\right)$. The two-component approach (generalized VMD plus hard pomeron) is used for a description of the underlying $\gamma^* N$-interaction. The hard component of the nucleon structure function is calculated in a framework of the colour dipole model with dipole cross section having Regge-type energy dependence. Numerical results for nuclear shadowing corrections are compared with available data of E665 and NMC collaborations.

## I. INTRODUCTION

This is a well-established fact that for small values of the Bjorken variable, $x \leq 0.1$, the nuclear structure functions of deep inelastic scattering, $F_{2A}(x,Q^2)$, are rather noticeably smaller than $A$ times the nucleon structure functions, $F_{2N}(x,Q^2)$, i.e., the corresponding virtual photon-nuclei cross sections are smaller than $A$ times the virtual photon-nucleon cross section. This is the well-known nuclear shadowing effect (see, e.g., [1-3]). The calculation of the nuclear shadowing in a broad interval of variables $x$ and $Q^2$ suffers from two main problems. Even if one uses a well elaborated method of the calculation (i.e., the method based on the $QCD$ perturbation theory) it fails in a region of small $Q^2$ $\left(Q^2 \leq 2-3\,GeV^2\right)$ because at these $Q^2$-values higher twist and nonperturbative effects are significant. For an accounting of these effects the contribution of vector mesons to nuclear shadowing, using the vector meson dominance (VMD) [4] model, must be calculated. The second problem is that the relative importance of these nonperturbative and higher twists effects, changing with a decrease of $x$, doesn't die out completely. Even in the region of extremely small $x$, $x \leq 10^{-9}$, they are noticeable at $Q^2 \sim 1\,GeV^2$.

In the present work we calculate nuclear shadowing functions for light nuclei, using the two-component (VMD + perturbative QCD) model elaborated in our previous articles [5-7]. In this approach the calculation of VMD component of DIS structure functions takes into account, except of $\rho$-meson, also its excited states ($\rho'$, $\rho''$,...) and ,what is essential, nondiagonal transitions between different members of this recurrence. It is clear that a calculation of the corresponding shadowing corrections is a complicate many-channel problem. Therefore, we were forced to use for diffractive structure functions (just these functions are needed for calculations of shadowing) phenomenological parameterizations based on experimental data of HERA. We suppose that all vector mesons of the model (except of the lightest ones which are considered separately) constitute together, effectively, the high-mass continuum which determines the diffractive cross section $d\sigma^{cont}/dM_X^2$. We assume, further, that diffractive DIS is a soft (nonperturbative) process (largely) [8] as well as inclusive DIS from the (generalized) VMD-component in the approach of [7]. In both cases the leading twist contributions to the structure functions are given by aligned configurations of $q\bar{q}$-pairs [9-11] produced by the virtual photon.

Calculating the shadowing corrections in a case of the perturbative $QCD$ component we use standard formulas of the eikonal approximation, in which a part of the total $q\bar{q}$-dipole-nucleon cross section, corresponding to the hard (perturbative) pomeron exchange, enters (here, as in the corresponding calculation of $F_{2N}$ in [7], the parameterization of the Regge-type $s$-dependence of the dipole cross section suggested in the FKS model [12] was used).

One should mention several works, in which the calculation of shadowing corrections had been carried out in a way which is closest to ours. First of all, it is the two-phase model including vector mesons and the (soft) pomeron [13]. The similar approach has been used in [14, 15] and in the more recent work [16].

The plan of the paper is as follows. In the second Section we calculate the shadowing correction to the nuclear structure function $F_{2A}$ taking into account only VMD component of $F_{2A}$. In the third Section the same correction is calculated, but only the perturbative $QCD$ component of $F_{2A}$ is considered. The last Section continues the comparison of the calculations with experimental data for two nuclei and our conclusions.

## II. VMD COMPONENT OF THE NUCLEAR STRUCTURE FUNCTION

As is said in the Introduction we approximate the spectrum of hadronic fluctuations of the virtual photon in VMD approach by one (or several) discrete vector meson masses plus continuum with border mass value $M_{X,\min}$. Correspondingly, the shadowing correction, which is defined by the relation

$$F_{2A}^{soft}\left(x,Q^2\right) = AF_{2N}^{soft} - \delta F_{2A}^{soft} \tag{1}$$

(upper index "soft" designates VMD component of the nuclear structure function), consists of two parts,

$$\delta F_{2A}^{soft} = \delta F_{2A}^{\rho,\rho',\ldots} + \delta F_{2A}^{cont}. \tag{2}$$

To determine the shadowing correction from the continuum part we need the diffractive structure function $F_2^{D(3)}$ which depends on three kinematic variables $Q^2, x_P, \beta$ connected with Bjorken variable $x$ and mass $M_X$ of the diffractive hadronic state by relations

$$\beta = \frac{Q^2}{Q^2 + M_X^2} = \frac{x}{x_P}. \tag{3}$$

In the literature there are many parameterizations of $F_2^{D(3)}$ obtained with using of HERA data. The most convenient for our aims (and most close to our VMD approach ideologically) is the parameterization suggested by W. Buchmuller and A. Hebecker [17]. Their fit is based on the observation [18] of the similarity between $x$ and $Q^2$ dependences of $F_2^{D(3)}$ and $F_{2N}$ in the small $x$ region (that is consistent with an assumption that diffractive and inclusive cross sections are both determined by the same dominant partonic process). The parameterization of [17] is

$$F_2^{D(3)}\left(x,Q^2,M_X^2\right) = \frac{0.04}{x_P} F_{2N}\left(x = x_P, Q^2\right). \tag{4}$$

In our case it is logical to use in this expression only the soft part of $F_{2N}$ (i.e., the VMD component of it),

$$F_{2N}^{soft} = F_{2N}^{total} - F_{2N}^{pert.QCD}. \tag{5}$$

In numerical calculations we used, for a comparison, also another fits suggested, in particular, by the concept of partonic pomeron [7]. For example, we used the fit [20, 21]

$$F_2^{D(3)}\left(x,Q^2,M_X^2\right) = \frac{1}{x_P^n} d\left[\beta(1-\beta) + \frac{f}{2}(1-\beta)^2\right], \tag{6}$$

$$n = 1.19,\ d = 0.03,\ f = 0.6.$$

At small values of $\beta$ the term in Eq. (6) which is proportional to $f$ and corresponds to a contribution of the soft pomeron exchange, dominates.

The contribution of the hadronic continuum to the shadowing correction is calculated using the formula [22,23]

$$\delta F_{2A}^{cont} = \frac{A(A-1)}{2} F_{2N}^{cont} \operatorname{Re}\left(\int d^2 b \int_{-\infty}^{\infty} dz_1 \int_{z_1}^{\infty} dz_2 (1-i\eta)^2 \sigma_{eff}\left(x,Q^2,z_2-z_1\right) \times \right. \\ \left. \rho_A(b,z_1)\rho_A(b,z_2) e^{-\frac{A}{2}(1-i\eta)\sigma_{eff}\left(x,Q^2,0\right)\int_{z_1}^{z_2}\rho_A(b,z)dz}\right). \tag{7}$$

Here, the cross section $\sigma_{eff}$ is given by the expression

$$\sigma_{eff}\left(x,Q^2,z_2-z_1\right) = \frac{16\pi B}{F_{2N}^{cont}(1+\eta^2)} \int_{x_{P\min}}^{x_{P\max}} dx_P F_2^{D(3)}\left(x,Q^2,M_X^2\right)\cos\left(\Delta_{M_X^2}(z_2-z_1)\right), \tag{8}$$

$$\Delta_{M_X^2} = xm_N\left(1+\frac{M_X^2}{Q^2}\right) = x_P m_N, \tag{9}$$

B is the exponential slope (assuming an exponential t-behavior of the diffractive cross section) (according to H1 data [24], B is ~ 5-6 GeV$^{-2}$), $\eta$ is the ratio of the real to imaginary parts of the diffractive scattering amplitude, $\rho_A(\vec{b},z)$ is the nuclear one-body density, normalized by the relation

$$2\pi\int_0^{\infty} b\,db \int_{-\infty}^{\infty} dz\,\rho_A(\vec{b},z) = 1. \tag{10}$$

The shadowing correction from the accounting of separate vector mesons (as a part of the spectrum of hadronic fluctuations of $\gamma^*$) is calculated using the formula

$$\delta F_{2A}^{\rho,\rho',...} = \frac{A(A-1)}{2} \frac{Q^2}{\pi F_{2N}^{soft}} \sum_{n=\rho,\rho',...} \left\{ \frac{e^2}{f_n^2} \eta_T \frac{M_n^4}{\left(Q^2+M_n^2\right)^2} \sigma_{n,T}^2 \times \right.$$

$$\int d^2 b \int_{-\infty}^{\infty} dz_1 \int_{z_1}^{\infty} dz_2 \rho_A(b,z_1) \rho_A(b,z_2) \cos(\Delta_n(z_2-z_1)) \exp\left[-\frac{A}{2}\sigma_{n,T}\int_{z_1}^{z_2}\rho_A(b,z)dz\right] +$$

$$\frac{e^2}{f_n^2} \eta_L \frac{M_n^4 Q^2}{\left(Q^2+M_n^2\right)^2} \sigma_{n,T}^2 \xi^2 \times$$

$$\left. \int d^2 b \int_{-\infty}^{\infty} dz_1 \int_{z_1}^{\infty} dz_2 \rho_A(b,z_1) \rho_A(b,z_2) \cos(\Delta_n(z_2-z_1)) \exp\left[-\frac{A}{2}\sigma_{n,T}\xi\int_{z_1}^{z_2}\rho_A(b,z)dz\right] \right\}, \quad (11)$$

$$\Delta_n = x m_N \left(1 + \frac{M_n^2}{Q^2}\right). \quad (12)$$

The equations of such type are derived in the standard (non-generalized) VMD model [25]; it has a form of the usual Glauber-Gribov [26] approximation for an interaction of the incoming virtual photon with two nucleons of the target nucleus, and the rescattering on these and more nucleons is included by introducing the attenuation factor. The similar quasi-eikonal approximation is used in deriving of Eq. (7), but now the rescattering cross section $\sigma_{eff}$ is equal to $\sigma_{V_n N} \equiv \sigma_n$, where $V_n$ is the vector meson with transverse (T) or longitudinal (L) polarization. The main specific feature of the aligned-jet version of the VMD model elaborated in [7] and used in this work is an introducing in Eq. (12) the $\eta_{T,L}$-factors which separate $q\bar{q}$-pairs (produced by $\gamma^*$) with large transverse size interacting with the target nonperturbatively. These factors lead to a decreasing of the effective $\gamma V$-coupling for heavy vector mesons in DIS and in Eq. (12),

$$\frac{e^2}{f^2} \eta_{T,L} \ll \frac{e^2}{f^2}. \quad (13)$$

The function $\xi(s)$ in Eq. (12) is the $\sigma_{n,L}/\sigma_{n,T}$-ratio. According to [7], it is approximately given by the expression

$$\xi(s) = \begin{cases} 0.25, & s \leq 30 \, GeV^2, \\ 0.17 \log s, & 30 \leq s \leq 7\,10^5 \, GeV^2 \\ 1, & s \geq 7\,10^5 \, GeV^2. \end{cases} \quad (14)$$

The number of vector mesons taken into account separately from the continuum (and the border mass value $M_{X,\min}$) can be determined from a comparison of a result of numerical calculations with shadowing data.

The total shadowing correction to $F_{2A}$ arising from taking into account of the VMD component of $F_{2N}$ is

$$\alpha_{soft} = \frac{F_{2A}^{soft}}{A F_{2N}^{soft}} = 1 - \frac{\delta F_{2A}^{\rho,\rho',...} + \delta F_{2A}^{cont}}{A F_{2N}^{soft}}. \quad (15)$$

### III. HARD COMPONENT OF THE NUCLEAR STRUCTURE FUNCTION

In the two-component model of the present authors [7] the hard component arises from the interactions of virtual $q\bar{q}$-pairs of non-aligned (symmetric) configurations with the nucleon target. These pairs have relatively small transverse sizes and their interaction with the nucleon can be described, in a language of Regge theory, by an exchange of the perturbative (hard) pomeron. For this description, in [7] the colour dipole model was used, in which the DIS structure functions and $\gamma^* N$-cross sections are given by the integrals

$$\sigma_{T,L}(Q^2,s) = \int dz \int d^2 r_\perp \left|\psi^{T,L}(r_\perp,z,Q^2,m_f)\right|^2 \sigma_{q\bar{q}N}(r_\perp,z,s). \quad (16)$$

Here, $\psi^{T,L}$ are the light-cone wave functions of the virtual photon, $\sigma_{q\bar{q}N}$ is the total cross section of the interaction of $q\bar{q}$-pair with the nucleon, $r_\perp$ is the transverse separation of particles of the pair, $z$ is the fraction of the incoming photon light-cone energy for one quark of the pair. For a phenomenological description of the hard interaction part of $\sigma_{q\bar{q}N}$ we used the parameterization suggested in FKS [12] model; their parameterization has a special functional form providing the maximum of the cross section at small $r_\perp$ ($r_\perp \sim 0.3-0.4$ fm), the exponential decrease in the limit of large $r_\perp$ and relatively fast growth of the cross section with $s$ ($\sim s^{0.4}$).

The shadowing correction to $F_{2A}^{hard}$, which is defined by the relation

$$F_{2A}^{hard}(x, Q^2) = A F_{2N}^{hard} - \delta F_{2A}^{hard}, \tag{17}$$

is calculated by the quasi-eikonal formula (see, e.g. [23])

$$\delta F_{2A}^{hard} = \frac{A(A-1)}{2} \frac{Q^2(1-x)}{4\pi^2 \alpha_{em}} \times$$

$$(2\pi)^2 \operatorname{Re}\left( \int_0^1 dz \int_0^\infty r_\perp dr_\perp \sum_f |\psi(r_\perp, z, Q^2, m_f)|^2 \int_0^\infty b\,db \int_{-\infty}^\infty dz_1 \int_{z_1}^\infty dz_2 (1-i\eta)^2 [\sigma_{hard}(r_\perp, s)]^2 \times \right. \tag{18}$$

$$\left. \rho_A(b, z_1) \rho_A(b, z_2) e^{i 2x M_N (z_1 - z_2)} e^{-\frac{A}{2}(1-i\eta)\sigma_{hard}(r_\perp, s)\int_{z_1}^{z_2} dz \rho_A(b,z)} \right).$$

The square of the virtual photon wavefunctions in this equation includes averaging over the photon helicities and is given by the expression ($K_0$ and $K_1$ are modified Hankel functions)

$$|\psi(r_\perp, z, Q^2, m_f)|^2 = \frac{6\alpha_{em}}{\pi^2} \sum_f \left(\frac{e_f}{e}\right)^2 \left[ \left( Q^2 z^2 (1-z)^2 + \frac{m_f^2}{4} \right) K_0^2(\bar{Q}_f r_\perp) + \right.$$

$$\left. \frac{1}{4}(z^2 + (1-z)^2) \bar{Q}_f^2 K_1^2(\bar{Q}_f r_\perp) \right], \tag{19}$$

$$\bar{Q}_f^2 = Q^2 z(1-z) + m_f^2, \tag{20}$$

$e_f$ and $m_f$ are charge and mass of the quark with the flavor $f$.

Finally, the shadowing correction for the hard component of $F_{2A}$ is

$$\alpha_{hard} = \frac{F_{2A}^{hard}}{A F_{2N}^{hard}} = 1 - \frac{\delta F_{2A}^{hard}}{A F_{2N}^{hard}}. \tag{21}$$

## IV. RESULTS OF CALCULATIONS AND CONCLUSIONS

All calculations in the present paper were carried out using the simplest assumption for a hadronic spectrum of virtual photon fluctuations: one separate vector meson ($\rho$) and continuum with the border mass 1.5 GeV. This choice gives the most satisfactory agreement with available data.

The total relative shadowing correction for $F_A$ in our two-component model is given by the sum

$$\alpha = \frac{F_{2A}^{soft} + F_{2A}^{hard}}{A(F_{2N}^{soft} + F_{2N}^{hard})} = \frac{F_{2N}^{soft}}{F_{2N}} \alpha_{soft} + \frac{F_{2N}^{hard}}{F_{2N}} \alpha_{hard}. \tag{22}$$

For the colour dipole cross section $\sigma_{hard}(r_\perp, s)$ we used the parameterization mentioned above,

$$\sigma_{hard}(r_\perp, s) = \left( \alpha_2^H r_\perp^2 + \alpha_6^H r_\perp^6 \right) e^{-\nu_H r_\perp} \left( r_\perp^2 s \right)^{\lambda_H}. \tag{23}$$

The values of parameters, which were taken from [12], are

$$\alpha_2^H = 0.072, \quad \alpha_6^H = 1.89, \quad \nu_H = 3.27, \quad \lambda_H = 0.44. \tag{24}$$

One must note that these values have been obtained by authors of [12] from fitting their total dipole cross section, $\sigma_{q\bar{q}N}(r_\perp, s)$, which has, except of the hard part, also the soft part, proportional to $s^{\lambda_s}$ ($\lambda_s \sim 0.06-0.08$). Our two-component model [7] describes the soft contribution to $F_{2N}$ by completely another way (by the generalized VMD approach). It appeared, nevertheless, that the fit of Eq. (22) is suited, with minimum corrections, for a description of the hard pomeron of the approach of [7]. In [7] the parameter $\nu_H$ slowly increases with a decrease of $x$ ($\nu_H = 3.27; 4; 5$ at, correspondingly, $x = 10^{-5}, 10^{-7}, 10^{-9}$).

For numerical calculations a two-parameter Fermi form of the nuclear one-body density was used:

$$\rho_A(\vec{b},z) = \frac{\rho_0}{1+\exp((r-c)/a)}, \quad r = \sqrt{|\vec{b}|^2 + z^2}, \quad a = 0.545. \tag{25}$$

For $^{40}$Ca $\rho_0 = 0.0039769\ fm^{-3}$, $c = 3.6663\ fm$, for $^{12}$C $\rho_0 = 0.013280\ fm^{-3}$, $c = 2.2486\ fm$.

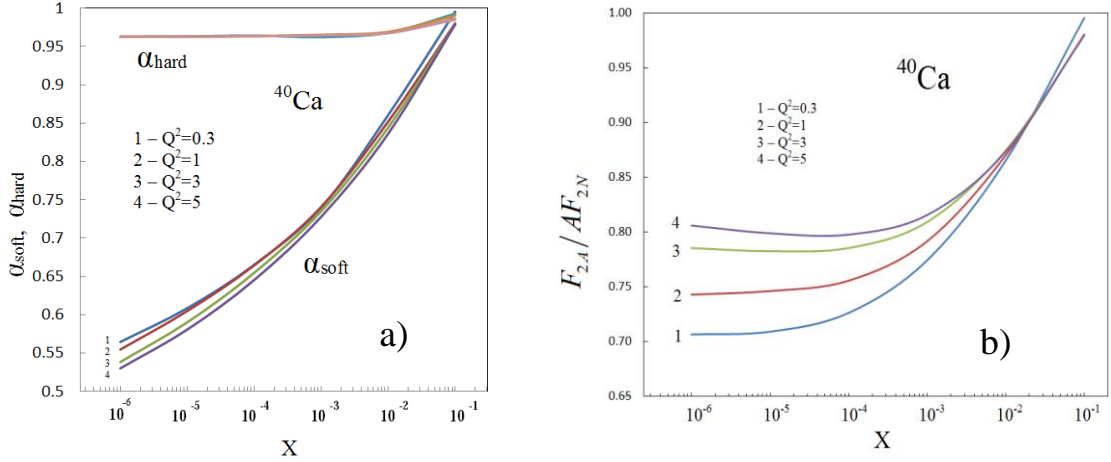

FIG. 1. Partial shadowing corrections defined in Eqs. (15) and (17) of the text (a) and total shadowing corrections (b) for $^{40}$Ca, as functions of $x$, for several values of $Q^2$.

The dependences of partial shadowing corrections, $\alpha_{soft}$ and $\alpha_{hard}$, on $x$, for several values of $Q^2$, are shown on fig.1a (for $^{40}$Ca nucleus). One can see from the figure the striking difference in behavior of the corrections as functions of $x$: the shadowing of the soft component is much stronger at $x < 10^{-2}$. The reason is very simple and is connected with a large contribution of diffractively produced inelastic states when large-size dipoles interact with the nucleon target. At the same time, the eikonal approximation (Eq. (18)) used for calculations with small-size dipoles neglects completely any diffractively produced inelastic states, such as $q\bar{q}g, q\bar{q}gg$ ... However, the corresponding underestimation of an amount of the shadowing in the case of small-size dipoles is essential only at large $Q^2$, in the region which we do not consider in the present paper.

On fig. 1b the total relative shadowing corrections as functions of $x$, are shown for several values of $Q^2$. It is clearly seen that the slopes of the curves change with $x$, in accordance, mostly, with a change of the relative fraction of the perturbative component in $F_{2N}$.

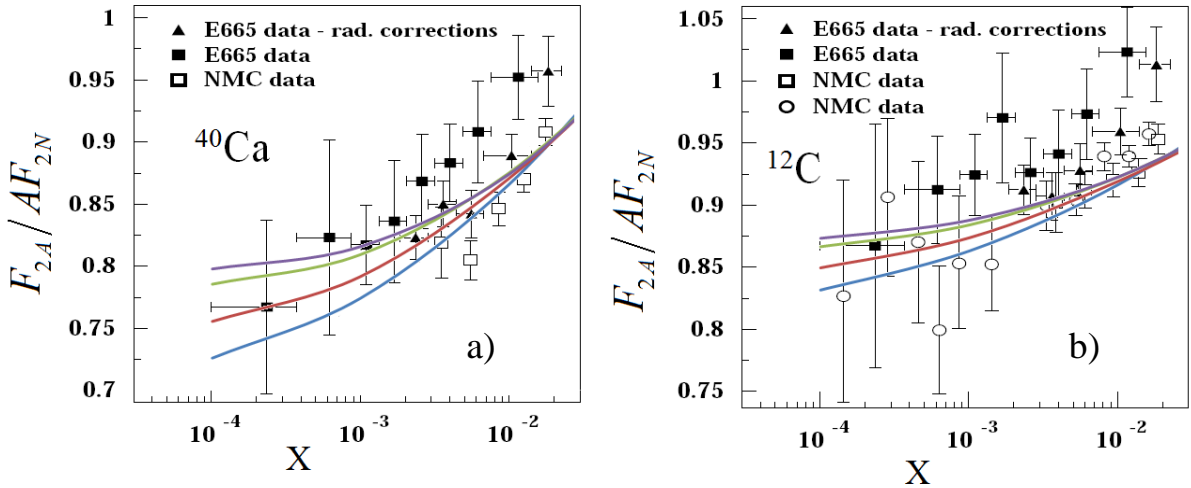

FIG. 2. The comparison of numerical results for $^{40}$Ca (a) and $^{12}$C (b) with data. Each curve corresponds to a fixed value of $Q^2$. From bottom to top: $Q^2 = 0.3, 1, 3, 5$.

The comparison of the numerical calculations with experimental data of NMC [27,28] and E655 [29,30] collaborations is shown on fig. 2 for two nuclei (the collection of all data points on the figure was taken from [31]). One can see that the agreement with NMC data is rather good at $x < 2\,10^{-2}$. One can see also that the calculated curves for the total $F_{2A}/AF_{2N}(x)$ show the gradual flattening with a decrease of $x$, for each given value of $Q^2$ as (likely) required by data. The character of the $Q^2$-dependence of the shadowing at fixed values of $x$ (in regions of low values of $x$) (larger $Q^2$ corresponds to a smaller shadowing) is similar with the predictions, e.g., of leading twist QCD approaches [23].

## ACKNOWLEDGEMENTS

We acknowledge the support of the Russian Federation Ministry of Education and Science (agreement 14.B25.31.0010 and government assignment 2014/51, project 1366, zadanie № 3.889.2014/K)